\documentclass[11pt,preprint]{emulateapj}
\usepackage{graphicx}

\begin{document}

\title{ALMA band~8 continuum emission from Orion Source~I}

\author{Tomoya Hirota\altaffilmark{1,2}, 
Masahiro N. Machida\altaffilmark{3}, Yuko Matsushita\altaffilmark{3}, 
Kazuhito Motogi\altaffilmark{4,5}, Naoko Matsumoto\altaffilmark{1,6}, 
Mi Kyoung Kim\altaffilmark{7}, Ross A. Burns\altaffilmark{8}, \& Mareki Honma\altaffilmark{4,9}} 
\email{tomoya.hirota@nao.ac.jp}
\altaffiltext{1}{Mizusawa VLBI Observatory, National Astronomical Observatory of Japan, Osawa 2-21-1, Mitaka, Tokyo 181-8588, Japan}
\altaffiltext{2}{Department of Astronomical Sciences, SOKENDAI (The Graduate University for Advanced Studies), Osawa 2-21-1, Mitaka, Tokyo 181-8588, Japan}
\altaffiltext{3}{Department of Earth and Planetary Sciences, Faculty of Sciences, Kyushu University, Motooka 744, Nishi-ku, Fukuoka, Fukuoka 819-0395, Japan}
\altaffiltext{4}{Mizusawa VLBI Observatory, National Astronomical Observatory of Japan, Hoshigaoka2-12, Mizusawa-ku, Oshu, Iwate 023-0861, Japan}
\altaffiltext{5}{Graduate School of Science and Engineering, Yamaguchi University, Yoshida 1677-1, Yamaguchi, Yamaguchi 753-8511, Japan}
\altaffiltext{6}{The Research Institute for Time Studies, Yamaguchi University, Yoshida 1677-1, Yamaguchi, Yamaguchi 753-8511, Japan}
\altaffiltext{7}{Korea Astronomy and Space Science Institute, Hwaam-dong 61-1, Yuseong-gu, Daejeon, 305-348, Republic of Korea}
\altaffiltext{8}{Joint Institute for VLBI in Europe, Postbus 2, 7990 AA, Dwingeloo, The Netherlands}
\altaffiltext{9}{Department of Astronomical Sciences, SOKENDAI (The Graduate University for Advanced Studies), Hoshigaoka2-12, Mizusawa-ku, Oshu, Iwate 023-0861, Japan}

\begin{abstract}
We have measured continuum flux densities of a high-mass protostar candidate, a radio source~I in the Orion~KL region (Orion Source~I) using the Atacama Large Millimeter/Submillimeter Array (ALMA) at band~8 with an angular resolution of 0.1\arcsec. 
The continuum emission at 430, 460, and 490~GHz associated with Source~I shows an elongated structure along the northwest-southeast direction perpendicular to the so-called low-velocity bipolar outflow. 
The deconvolved size of the continuum source, 90~au$\times$20~au, is consistent with those reported previously at other millimeter/submillimeter wavelength. 
The flux density can be well fitted to the optically thick black-body spectral energy distribution (SED), and the brightness temperature is evaluated to be 700-800~K. 
It is much lower than that in the case of proton-electron or H$^{-}$ free-free radiations. 
Our data are consistent with the latest ALMA results by \citet{plambeck2016}, in which the continuum emission have been proposed to arise from the edge-on circumstellar disk via thermal dust emission, unless the continuum source consists of an unresolved structure with the smaller beam filling factor. 
\end{abstract}


\section{Introduction}

The Orion Kleinmann-Low \citep[KL;][]{kleinmann1967} region is known to be the nearest site of high-mass star-formation at a distance of 420~pc from the Sun \citep{menten2007,kim2008}. 
There are numbers of deeply embedded young stellar objects (YSOs) detected in X-ray, infrared, and radio wavelengths. 
In particular, high resolution radio interferometer observations have identified dust continuum sources, H{\sc{ii}} regions, and radio jets driven by YSOs \citep[e.g.][]{garay1987, zapata2004, rivilla2015, forbrich2016}. 

Among a number of high-mass YSO candidates, a radio source~I (Orion Source~I) has the most outstanding characteristics. 
It has been thought to be a prototypical high-mass protostar candidate with a luminosity greater than 10$^{4}L_{\odot}$ \citep{menten1995}. 
Source~I is a driving source of a so-called low velocity (18~km~s$^{-1}$) bipolar outflow along the northeast-southwest (NE-SW) direction with a scale of 1000~AU traced by the thermal SiO lines \citep{plambeck2009,zapata2012,niederhofer2012,greenhill2013}. 
At the position of the central YSO, vibrationally excited SiO masers are detected, which is a quite rare case for star-forming regions \citep{zapata2009}. 
High resolution VLBI observations reveal that the SiO masers trace a magneto-centrifugal disk wind from the surface of the circumstellar disk with 100~au scale \citep{matthews2010}. 
Vibrationally excited molecular lines such as H$_{2}$O, SiO, CO, and SO at the excitation energies of 500-3500~K are also detected around Source~I by the recent Atacama Large Millimeter/Submillimeter Array (ALMA) observations suggesting a hot molecular gas in the rotating disk and base of the outflow \citep{hirota2012, hirota2014, hirota2016, plambeck2016}. 

Continuum emissions are detected from centimeter to submillimeter wavelengths up to 690~GHz \citep[][for the latest review]{plambeck2016}. 
Recent high resolution observations with the Very Large Array (VLA) and ALMA resolve the structure of the continuum emission with the deconvolved size of $\sim$100~au scale \citep{reid2007, goddi2011, plambeck2013, plambeck2016}. 
Because the continuum source is elongated perpendicular to the low-velocity NE-SW outflow, the continuum emission is thought to trace an edge-on  circumstellar disk rather than a radio jet. 
Possible mechanism of the continuum emission is proposed such as thermal dust gray-body radiation, H$^{-}$ free-free radiation, proton-electron free-free radiation, and their combination \citep{beuther2004, beuther2006, reid2007, plambeck2013, hirota2015, plambeck2016}. 
The spectral energy distribution (SED) of Source~I shows a power-law function with an index of 2 below 350~GHz while it becomes slightly steeper at higher frequency than 350~GHz \citep{plambeck2016}. 
Such a characteristics along with the lack of turnover in the SED \citep{hirota2015} strongly implies the optically thick thermal dust emission \citep{plambeck2016}. 
The brightness temperature of 500~K in the submillimeter bands also support an explanation via the thermal dust emission \citep{plambeck2016}. 

In this letter, we present the new continuum measurement results with ALMA at band~8 (430~GHz, 460~GHz, and 490~GHz) at a resolution of 100~mas or 40~au. 
Our data achieve the highest spatial resolution in the recent millimeter/submillimeter observations except that from VLA \citep{reid2007, goddi2011}, and hence, they are helpful to confirm the recent observational results \citep{hirota2015, plambeck2016}. 
Accurate measurements of the continuum flux densities can also constrain a shape of the SED such as a possible signature of excess emission and turnover frequency \citep[e.g.][]{beuther2004, beuther2006, plambeck2013, hirota2015}. 
Our results support the conclusions obtained from the latest ALMA observations at 350~GHz and 660~GHz \citep{plambeck2016}. 
The present newer and higher resolution data could put better constraints on the width of the minor axis, and establish an improved lower limit on the brightness temperature. 

\section{Observations and Data Analysis}

Observations were carried out in 2015 with ALMA as one of the science projects in cycle~2 (2013.1.00048). 
The tracking center position was RA(J2000)=05h35m14.512s, Decl(J2000)=-05d22'30.57". 
Details of observations are listed in Table \ref{tab-obs}. 
A primary flux calibrator and band-pass calibrator were J0423-013, and J0522-3627, respectively. 
Secondary gain calibrators were J0607-0834 for observations at 460~GHz and 490~GHz and J0501-0159 at 430~GHz. 
The ALMA correlator provided the four spectral windows with the total bandwidth of 
1875~MHz and 937.5~MHz for two windows each. 
The channel spacings of spectrometers were 976.562~kHz and 488.281~kHz for the spectral windows with the 1875~MHz and 937.5~MHz bandwidths, respectively. 
Dual polarization data were obtained simultaneously for all the frequency bands. 

The data were calibrated and imaged with the Common Astronomy Software Applications (CASA) package. 
We employed the calibrated data delivered by the East-Asia ALMA Regional Center (EA-ARC). 
First, visibility data were separated into spectral lines and continuum emissions by setting the line free channels with the CASA task {\tt{uvcontsub}}. 
Next, both phase and amplitude self-calibration was done with the continuum emission of Source~I by integrating over all the channels using CASA tasks {\tt{clean}} and {\tt{gaincal}}. 
Because of calibration problems in the delivered data due to the large atmospheric opacity around 487~GHz, we used only three spectral windows among four ALMA basebands to make a continuum image at 490~GHz. 
For the intensity maps of the continuum emissions, we employed the visibility data with the UV distance longer than 400~k$\lambda$ to resolve out extended emission components \citep{hirota2015}. 
The procedure provides an angular resolution of $\sim$100~mas, corresponding to the linear resolution of 40~au at the distance to Orion~KL. 
We note that the observed peak positions of the continuum emissions at different frequencies are offset from each others as summarized in Table \ref{tab-cont}. 
It is probably due to astrometric calibration errors in the ALMA observations/data analysis. 
Thus, all the maps are registered based on the continuum peak positions, and we will not discuss about absolute position of each emission. 

\section{Results and discussion}

\begin{figure*}
\begin{center}
\includegraphics[width=15cm]{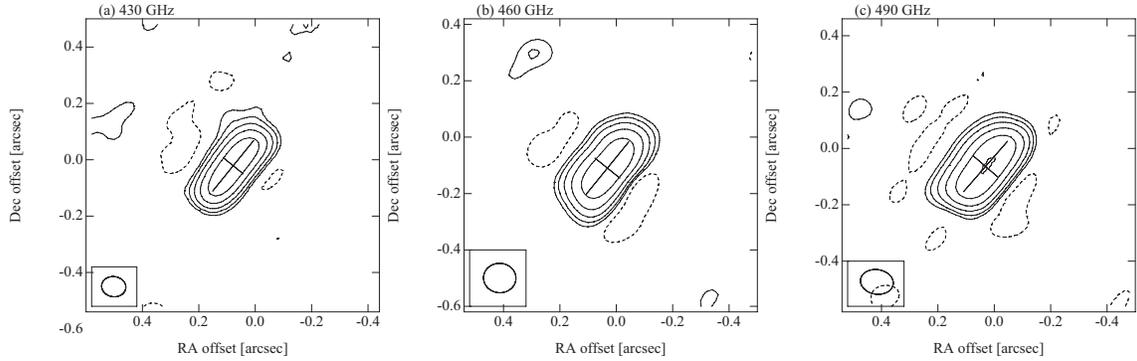}
\caption{
Continuum emission maps for band~8 data. 
Synthesized beam size is indicated at the bottom-left corner in each panel. 
The contour levels are $-$4, 4, 8, 16, 32, ... $\times$ the rms noise level. 
The (0,0) position is the tracking center position, RA(J2000)=05h35m14.512s, Decl(J2000)=$-$05d22\arcmin30.57\arcsec. 
(a) 430~GHz. The noise level (1$\sigma$) and peak intensity are 2.7~mJy~beam$^{-1}$ and 324~mJy~beam$^{-1}$, respectively. 
(b) 460~GHz. The noise level (1$\sigma$) and peak intensity are 4.1~mJy~beam$^{-1}$ and 481~mJy~beam$^{-1}$, respectively. 
(c) 490~GHz. The noise level (1$\sigma$) and peak intensity are 3.8~mJy~beam$^{-1}$ and 506~mJy~beam$^{-1}$, respectively. 
A cross in each panel represents the major and minor axis of the image. 
Note that the positions are offset from each other (see text). 
}
\label{fig-map}
\end{center}
\end{figure*}

Figures \ref{fig-map} shows continuum images of Source~I at three observed frequencies in ALMA band~8. 
The source properties are determined by fitting the two dimensional Gaussian to the images and the derived parameters are listed in Table \ref{tab-cont}. 
The continuum maps have the sizes of about (230-240)~mas$\times$(90-110)~mas with the position angle of 140~degrees. 
The structure is elongated along the NW-SE direction, which is perpendicular to the low-velocity NE-SW bipolar outflow. 
The deconvolved size is 220~mas$\times$50~mas, corresponding to 90~au$\times$20~au at the distance of Orion~KL. 
All the results are consistent with those derived from previous ALMA observations at 350~GHz and 660~GHz with slightly larger beam size \citep{plambeck2016}. 

\begin{figure*}
\begin{center}
\includegraphics[width=16cm]{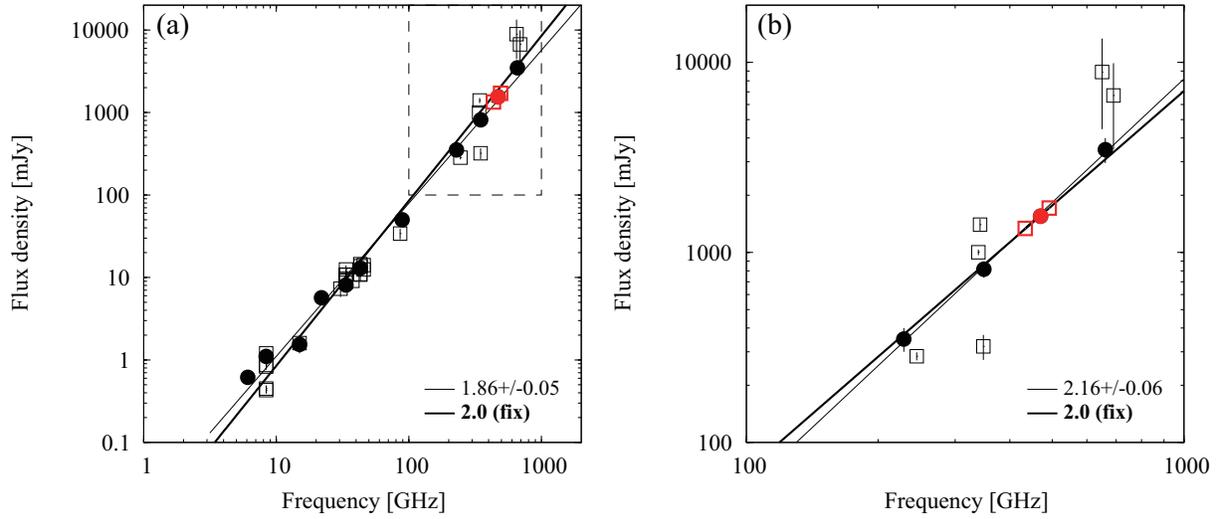}
\caption{
Spectral energy distribution (SED) of Source~I.
(a) All the frequency bands from 6 to 660~GHz. 
(b) Frequency bands enlarged from 100-1000~GHz as indicated by a dashed box in panel (a). 
Bold solid lines indicate the black-body SED with the fixed power-law index of 2.0. 
Thin solid lines show the best-fit single power-law models $F_{\nu}=p \nu^{q}$. 
The derived power-law indexes are shown in each panel. 
Red open and filled circles show our ALMA band~8 data at 430~GHz, 460~GHz, and 490~GHz. 
Black open squares and filled circles represent all the data including the present ALMA results \citep[][and references therein]{plambeck2013,hirota2015,plambeck2016}. 
In the SED fitting, we only employed the flux density data obtained with the highest spatial resolution in each band, as indicated by filled black and red circles. 
}
\label{fig-sed}
\end{center}
\end{figure*}

The flux densities are plotted in Figure \ref{fig-sed} along with previous observations \cite[][and references therein]{plambeck2013, hirota2015, plambeck2016}. 
Recent ALMA observations of continuum emissions at 350~GHz and 660~GHz show the optically thick spectral energy distribution (SED) with a power-law index of 2.0 \citep{plambeck2016}. 
Our measured flux densities agrees well with those of their interpolation. 

Because we used only visibilities with the longer baseline length than 400~k$\lambda$ in the imaging, some of the emission could be resolved out. 
Using the CASA task {\tt{simobserve}}, we simulated the ALMA imaging of a Gaussian source model with the size of 220~mas$\times$50~mas and total flux density of 1~Jy. 
When we assume the same uv coverage of our observation at 460~GHz with a uniform weighting of only $>$400~k$\lambda$ data, the derived flux density and deconvolved source size, 0.9996~Jy and 220.0~mas$\times$49.8~mas can reconstruct the model parameters. 
Thus, our imaging results provide reliable source size and flux density. 

We reanalyze the SED by employing our new data at 460~GHz band. 
The same fitting procedures are applied to the present data as described in \citet{hirota2015}. 
If there are more than two observed results previously reported at the same frequency band, we only employ the flux density data with the highest spatial resolution at each band. 
For our ALMA band~8 results, we only use the 460~GHz data and do not include the 430~GHz and 490~GHz data in order not to put too much weight for our band~8 data in the fitting. 
When all the data from 6 to 660~GHz are used in the fitting as shown in Figure \ref{fig-sed}(a), the best fit power-law SED, $F_{\nu}=p \nu^{q}$, has an spectral index $q$ of 1.86$\pm$0.05. 
Compared with the best fit SED, the black-body SED with the fixed index of 2.0 shows significant deviation at the lower frequency data below 10~GHz because of the excess emission from the free-free radiation as proposed by \citet{plambeck2016}. 
Alternatively, it may also suggest an unresolved spatial structure of inhomogeneous gas distribution or physical properties such as density/temperature distribution \citep{beuther2004, beuther2006, plambeck2013}. 

For comparison, we plot the SED at the frequency higher than 100~GHz in Figure \ref{fig-sed}(b). 
We cannot see a clear signature of a turnover in the SED that would suggest an optically thin H$^{-}$ free-free or proton-electron free-free radiation \citep{beuther2004, beuther2006, hirota2015}. 
The best-fit SED model by using only the high frequencies from 230~GHz to 660~GHz gives the power-law index of 2.16$\pm$0.06. 
In all cases, the SED at our band~8 data is consistent with the optically thick black-body radiation. 
We note that the best-fit power-law index of 2.16 is marginally larger than 2.0. 
This is consistent with the excess flux at higher frequency than 600~GHz where the dust emission has the larger source size or becomes  hotter as already reported by \citet{plambeck2016}. 

Because the source structure is marginally resolved in the present observations, we can derive the brightness temperature of the continuum emission, as listed in Table \ref{tab-cont}. 
The brightness temperature of 700-800~K is slightly higher than that derived from the 350~GHz and 660~GHz continuum data of 500~K \citep{plambeck2016}. 
It is bacause our higher resolution data can better constrain the source size, in particular for the minor axis of the edge-on disk. 
According to \citet{plambeck2016}, proton-electron free-free emission is unlikely to be an opacity source because of the lack of hydrogen recombination line and extremely large luminosity required for the optically thick SED. 
The optically thick H$^{-}$ free-free radiation is also ruled out since unrealistically high density or large disk mass is necessary to satisfy the optically thick SED without a turnover \citep{plambeck2016}. 
Our new SED fitting with the beam averaged brightness temperature of 700-800~K agrees with their interpretation via the optically thick thermal dust emission  \citep{plambeck2016}. 

Nevertheless, there may be a smaller scale structure unresolved with the ALMA beam $\sim$40~au which could have a higher brightness temperature. 
In fact, the higher spatial resolution data at 43~GHz observed with VLA show the more compact structure than the present ALMA beam size with the higher brightness temperature of 1600~K \citep{reid2007}. 
The excess flux densities below 43~GHz is attributed to a contribution from the hotter free-free radiation \citep{plambeck2016}. 
If the actual disk size or the beam filling factor is smaller by a factor of two than that of the deconvolved size, the brightness temperature becomes higher than 1600~K observed with the VLA. 
In this case, we cannot fully rule out a possibility of the H$^{-}$ free-free radiation as an opacity source of the emission. 
The hydrogen density to achieve the optically thick H$^{-}$ free-free emission up to 660~GHz requires $\sim$10$^{12}$-10$^{13}$~cm$^{-3}$ corresponding to a disk mass of 2$M_{\odot}$ \citep[e.g.][]{reid2007}. 
The highest resolution imaging that can be achieved with ALMA $<$10~au will be able to resolve a vertical and internal clumpy structure of this edge-on disk for the first time, which will be the key issue to understanding of the basic nature of Source~I. 

\acknowledgements
We would like to thank Richard L. Plambeck for valuable comments. 
This paper makes use of the following ALMA data: ADS/JAO.ALMA\#2013.1.00048.S. 
ALMA is a partnership of ESO (representing its member states), NSF (USA) and NINS (Japan), together with NRC (Canada), NSC and ASIAA (Taiwan), and KASI (Republic of Korea), in cooperation with the Republic of Chile. The Joint ALMA Observatory is operated by ESO, AUI/NRAO and NAOJ. 
We thank the staff at ALMA for making observations and reducing the science verification data. 
T.H. is supported by the MEXT/JSPS KAKENHI Grant Numbers 21224002, 24684011, 25108005, and 15H03646, and the ALMA Japan Research Grant of NAOJ Chile Observatory, NAOJ-ALMA-0006, 0028, 0066. 
M.H. is supported by the MEXT/JSPS KAKENHI Grant Numbers 24540242 and 25120007. 
Data analysis were in part carried out on common use data analysis computer system at the Astronomy Data Center, ADC, of the National Astronomical Observatory of Japan. 

{\it Facilities:} \facility{ALMA}.

{}

\begin{deluxetable*}{clcccrrcrc}
\tablewidth{0pt}
\tabletypesize{\scriptsize}
\tablecaption{Summary of Observations
 \label{tab-obs}}
\tablehead{
\colhead{Center}      & \colhead{} & \colhead{Total} & 
                                \colhead{Number} & \colhead{Longest} & \colhead{On-source} & \colhead{Median} & \colhead{beam size} & \colhead{} & \colhead{} \\
\colhead{Frequency\tablenotemark{a}} & \colhead{Date} & \colhead{bandwidth} & 
                                \colhead{of}     & \colhead{Baseline}     & \colhead{Time} & \colhead{$T_{sys}$} & \colhead{FWHM} & \colhead{PA} & \colhead{rms} \\
\colhead{(GHz)}       & \colhead{(in 2015)} & \colhead{(MHz)} & 
                                \colhead{Antennas} & \colhead{(m)} & \colhead{(sec)} & \colhead{(K)} & \colhead{(mas $\times$ mas)} & \colhead{(degree)} & \colhead{(mJy~beam$^{-1}$)}
}
\startdata
430 & Sep. 22 & 5625 & 35 & 2270 & 1929 & 1200 &  83$\times$67 & $-$79.8 & 2.7 \\
460 & Aug. 27 & 5625 & 40 & 1574 &  410 &   430 & 104$\times$87 &  82.9 & 4.1 \\
490 & Jul. 27 & 4688 & 41 & 1466 &  315 &   400 & 111$\times$79 &  84.5 & 3.8 
\enddata
\tablenotetext{a}{Center frequencies for LSB (lower side band) and USB (upper side band). }
\end{deluxetable*}

\begin{deluxetable*}{ccccrcrccc}
\tablewidth{0pt}
\tabletypesize{\scriptsize}
\tablecaption{Gauss fitting results 
\label{tab-cont}}
\tablehead{
\colhead{Frequency}                 & 
\colhead{$\Delta \alpha$\tablenotemark{a}}  & \colhead{$\Delta \delta$\tablenotemark{a}}  &
\colhead{Convolved size}   & \colhead{PA}   & 
\colhead{Deconvolved size}   & \colhead{PA}   & 
\colhead{Peak intensity}       & \colhead{Integrated flux} & \colhead{$T_{b}$} \\
\colhead{(GHz)}                              &
\colhead{(mas)}                     & \colhead{(mas)}         & 
\colhead{(mas $\times$ mas)}      & \colhead{(deg)}                           &
\colhead{(mas $\times$ mas)}      & \colhead{(deg)}                           &
\colhead{(mJy~beam$^{-1}$)}                           & \colhead{(mJy)}  & \colhead{(K)}
}
\startdata
440 &    78.5(5) & $-$21.5(5)  &  229.6(16)$\times$91.4(6)   & 141.4(3) & 216.7(16)$\times$53.0(12) & 143.1(3) &  353(2) & 1338(11) & 735(6) \\
460 &    31.7(5) & $-$109.7(6) & 238.8(16)$\times$112.6(8) & 141.3(3) & 220.4(18)$\times$52.3(18) & 143.1(4) &  520(4) & 1547(14) & 777(7) \\
490 &    31.0(4) & $-$64.0(5)  &  236.7(13)$\times$113.9(6) & 138.8(3) & 218.9(15)$\times$50.9(17) & 142.4(3) &  557(3) & 1709(12) & 756(6) 
\enddata
\tablecomments{Numbers in parenthesis represent fitting errors in unit of the last significant digits.}
\tablenotetext{a}{Offset from the tracking center position. }
\vspace{5mm}
\end{deluxetable*}

\end{document}